\documentclass[multphys,vecphys]{svmult}

\usepackage{makeidx}         
\usepackage{graphicx}        
\usepackage{multicol}        
\usepackage[bottom]{footmisc}

\makeindex             
\newcommand{\ds}{\displaystyle}

\begin{document}

\title*{High-Precision Spectroscopy of Pulsating Stars}

\author{C.\ Aerts\inst{1,2}\and S.\ Hekker\inst{1,3}\and M.\ Desmet\inst{1}
\and F.\ Carrier\inst{1} \and W.\ Zima\inst{1} \and M.\ Briquet\inst{1}
\and J.\ De Ridder\inst{1}}

\institute{
Institute of Astronomy, Catholic University of Leuven, Celestijnenlaan 200D,
B-3001 Leuven, Belgium
\texttt{conny@ster.kuleuven.be}
\and Department of Astrophysics, Radboud University Nijmegen,
 P.O. Box 9010, 6500 GL Nijmegen, The Netherlands
\and
Leiden Observatory, P.O. Box 9513, 2333 RA Leiden, The Netherlands}
%
%
\maketitle

\begin{abstract}
We review the methodology currently available to interprete time series of
high-resolution high-S/N spectroscopic data of pulsating stars in terms of the
kind of (non-radial) modes that are excited. We illustrate the drastic
improvement of the detection treshold of line-profile variability thanks to the
advancement of the instrumentation over the past two decades.  This has led to
the opportunity to interprete line-profile variations with amplitudes of order
m/s, which is a factor 1000 lower than the earliest line-profile time series
studies allowed for.
\end{abstract}

\section{Line-Profile Variations due to Pulsations}

In the recent research domain of asteroseismology, one tries to probe the
internal structure parameter of stars from their observed pulsation properties.
Prerequisites to succeed in that are accurate pulsation frequency values and an
unambiguous identification of the spherical wavenumbers $(\ell,m)$ of at least
two, but preferrably many more, non-radial pulsation modes. Asteroseismology has
been applied successfully across the whole HR diagram. For a recent extensive
overview of its successes so far, we refer to Kurtz (2006).

\begin{figure}[t]\begin{center}
\rotatebox{0}{\resizebox{7.5cm}{!}{\includegraphics{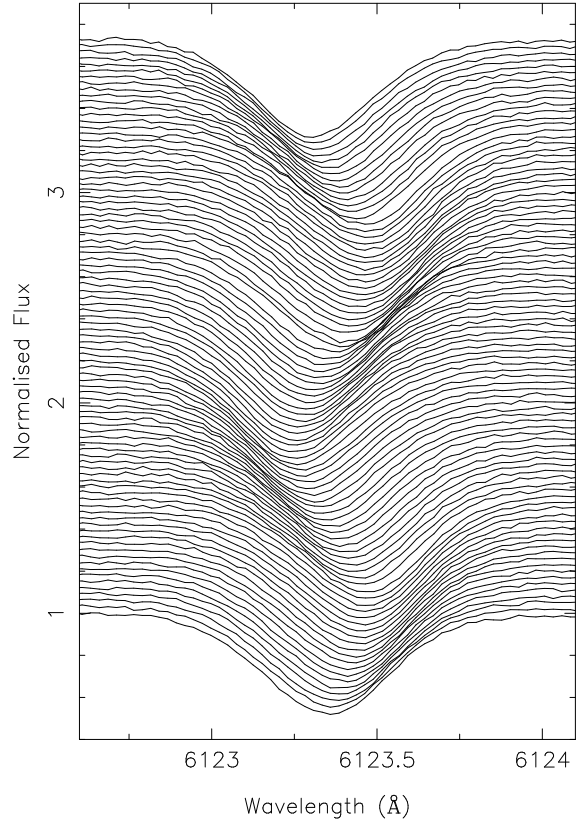}}}
\end{center}
\caption{Observed line-profile variations of the $\delta\,$Sct star
$\rho\,$Puppis obtained in 1995 with the Coud\'e Auxiliary Telescope of the
European Southern Observatory in Chile.  Data taken from Mathias et al.\
(1997). }
\label{rpup}
\end{figure}

The introduction of high-resolution spectrographs with sensitive detectors in
the 1980s had a large impact on the field of pulsation mode identification.
Spectroscopic data indeed offer a very detailed picture of the pulsation
velocity field. Mode identification requires that moderate to large
telescopes be available during a long time span. Indeed, it remains a
challenge to obtain time-resolved spectra with a high resolving power and with a
high signal-to-noise ratio ($> 300$) covering the overall beat period of the
multiperiodic pulsation. The required temporal resolution must be such that the
ratio of the integration time and the pulsation period(s) remains below a few
percent. The latter condition is necessary in order to avoid smearing out the
effects of the pulsations in the line profiles during the cycle. This
requirement is difficult to meet for rapid pulsators with periods of order
minutes. This is the reason why spectroscopic mode identification has been
achieved mainly for massive main sequence stars with spectral type from O to F
and with pulsation modes excited by the $\kappa\,$mechanism with periods above
one hour. An example of a nice data set is shown in Fig.\,\ref{rpup}. The
application to stars with stochastic excitation, to magnetic pulsators and to
compact pulsators has only recently been attempted and needs further
improvements.

\section{Methodology for Line-Profile Analysis}

The pulsation velocity due to a spheroidal pulsation mode with infinite lifetime
in the approximation of a non-rotating star equals
\begin{equation}
\label{nrp}
\ds{\vec{v}_{\rm puls}=
\left(v_r,v_{\theta},v_{\varphi}\right)=N_{\ell}^mv_{\rm p}
\left(1,K\frac{\partial}{\partial\theta},\frac{K}{\sin\theta}
\frac{\partial}{\partial\varphi}\right)Y_{\ell}^m(\theta,\varphi)
\exp{\left({\rm i}\omega t\right),}}
\end{equation}
when a system of spherical coordinates $(r,\theta,\varphi)$ with origin at the
centre of the star and with polar axis along the rotation axis is used. In this
expression, $N_{\ell}^m$ is a normalisation factor for the
$Y_{\ell}^m(\theta,\varphi)$, $v_{\rm p}$ is proportional to the pulsation
amplitude, $\omega$ is the cyclic pulsation frequency, and $K$ is the ratio of
the horizontal to the vertical velocity amplitude: $K=GM/(\omega^2R^3)$.  This
velocity acts together with rotational broadening and with intrinsic line
broadening due to pressure and temperature effects. Moreover, we only detect the
velocity component projected onto the line-of-sight. This leads us to the
conclusion that line-profile variations due to a single pulsation mode are
determiend by six unknown parameters among which two are integer numbers
$(\ell,m)$.  For each additional mode, three unknowns are added in the linear
approximation ignoring non-linear coupling between the modes.  The above
Exp.\,(\ref{nrp}) is far too simple when Coriolis, centrifugal or Lorentz forces
come into play.

It is clear from these arguments that the derivation of the full
details of the pulsational velocity field from observed line-profile variations
cannot be a simple task. Nevertheless, the richness of the information in these
type of data is such an asset compared to photometric data (which essentially
only allow estimation of $\ell$) that spectroscopic mode identification has
become an entire subfield by itself within asteroseismology.

Fairly recent overviews of the methodology for spectroscopic mode
identification, ideally suited for the unexperienced reader, are available in
Telting \& Schrijvers (1997), Aerts \& Eyer (2000), and Mantegazza
(2000). Rather than repeating what is available in these papers, we point out
two newer versions of the methods available since then. One is the numerical
implementation of the so-called moment method (Briquet \& Aerts 2003).  In this
work, the authors have generalised previous versions of this technique, which is
based on the time-variations of the lowest-order moments of a line profile and
which works well for slow rotators, to multiperiodic pulsators. It has meanwhile
been applied to several stars whose pulsational broadening dominates over
rotational and intrinsic broadening. Zima (2006), on the other hand, generalised
the Telting \& Schrijvers (1997) method, which is based on the amplitude and
phase variations of the modes across the line profile, to include a statistical
significance criterion for the mode identification and applied it to the
rapidly-rotating $\delta\,$Sct star FG\,Vir (Zima et al.\ 2006). A recent
combined application of both methods to the pulsations of the $\beta\,$Cep star
$\nu\,$Eri, observed during a multisite campaign, was made by De Ridder et al.\
(2004), while De Cat et al.\ (2005) applied both techniques to the observed
line-profile variations of several slowly pulsating B stars.

In all of the abovementioned examples, one carefully selected isolated and
unblended spectral line was used for the analysis. This is a good strategy
whenever pulsation amplitudes above a few km/s are encountered and when the
required temporal resolution is achieved.  For lower-amplitude pulsators,
however, or when time-resolved spectroscopy leads to a too low signal-to-noise
ratio, one may also apply the methodology discussed above to a time series of
cross-correlation functions (CCFs) derived from different lines in the
spectrum. This induces complications, however, because one takes a weighted
average over a much more extended line-forming region in the stellar atmosphere
compared to the case where one uses only one line. The pulsation amplitude and
phase may have slightly different behaviour across this whole region, such that
one models their average value in that case. This works fine on the condition
that no nodal surfaces of the pulsation eigenfunction occur in the line-forming
region. An adaptation of the moment method in such a case was already made by
Mathias \& Aerts (1996), who applied it to the low-amplitude ($< 5\,$km/s)
$\delta\,$Sct star 20\,CVn. More recently, Aerts et al.\ (2004) and De Cat et
al.\ (2006) also used CCFs to analyse extensive spectroscopic data of a sample
of $\gamma\,$Dor stars, with pulsation amplitudes typically below 2\,km/s.

\section{Towards lower amplitudes}

The first and so far only application of the line-profile methodology to
solar-like oscillations with finite lifetimes was made by Hekker et al.\
(2006). They used extensive time series of CCFs of four pulsating red giants
derived from CORALIE spectra obtained with the 1.2m Euler telescope in an
attempt to identify the pulsation modes. Their results are based on a simulation
study in which they assessed the effect of the finite mode lifetime on the
line-profile diagnostics, resulting in the conclusion that the phase across the
CCF can no longer be used for mode identification.  Due to this, and the very
low amplitudes of $\sim$m/s, only an estimate of $(\ell,m)$ of the dominant mode
could be obtained. This led to the surprising result that these stars seem to
have non-radial modes, while it was assumed so far in the modelling that the
frequencies were due to radial modes. The application to solar-like pulsators is
of relevance for exoplanet search as well, since a Keplerian radial-velocity
shift is then superimposed on the pulsation velocity. It remains to be studied
how well one can separate between those two effects at such low amplitudes.

%
%
%

%
%



\printindex
\end{document}